\documentclass[11pt,prd,nofootinbib,reprint,superscriptaddress,longbibliography]{revtex4-2}
\usepackage{amsmath, amssymb, amsthm, graphicx, epsfig, fancyhdr,epsfig,multirow}
\usepackage[utf8]{inputenc}
\usepackage{amsmath}
\usepackage{amsfonts}
\usepackage{amssymb}
\usepackage{xcolor}
\usepackage{comment}
\usepackage{subcaption}
\usepackage[normalem]{ulem}
\usepackage[left=2cm,right=2cm,top=2cm,bottom=2cm]{geometry}
\usepackage[font=small,labelfont=bf,textfont=it]{caption}
\usepackage[toc,page]{appendix}

\DeclareUnicodeCharacter{2009}{\,} 
\DeclareUnicodeCharacter{2212}{-}
\newcommand{\nn}{\nonumber}
\newcommand{\xenon}{{\sc{Xenon}}n{\sc{T}}}
\newcommand{\superkam}{\sc Super-Kamiokande}
\newcommand{\superk}{\sc Super-K}
\usepackage{color}
 
\begin{document}
\title{Bounds on boosted dark matter from direct detection: \\
The role of energy-dependent cross sections}
\author{Debjyoti Bardhan} 
\email{debjyoti.bardhan@acads.iiserpune.ac.in}
\affiliation{Department of Physics, Indian Institute of Science Education and Research Pune, India}
\author{Supritha Bhowmick}
\email{supritha.bhowmick@students.iiserpune.ac.in}
\affiliation{Department of Physics, Indian Institute of Science Education and Research Pune, India}
\author{Diptimoy Ghosh}  
\email{diptimoy.ghosh@iiserpune.ac.in}
\affiliation{Department of Physics, Indian Institute of Science Education and Research Pune, India}
\author{Atanu Guha}
\affiliation{Department of Physics, Indian Institute of Science Education and Research Pune, India}
\affiliation{Department of Physics, Chungnam National University, South Korea}
\email{atanu@cnu.ac.kr}
\author{Divya Sachdeva}
\email{dsachdeva@lpthe.jussieu.fr}
\affiliation{Laboratoire de Physique Th\'eorique et Hautes \'Energies (LPTHE), UMR 7589 CNRS and Sorbonne Universit\'e, 4 Place Jussieu, F-75252, Paris, France}

\begin{abstract}
The recoil threshold of Direct Detection (DD) experiments limits the mass range of Dark Matter (DM) particles that can be detected, with most DD experiments being blind to sub-MeV DM particles. However, these light DM particles can be boosted to very high energies via collisions with energetic Cosmic Ray electrons. This allows Dark Matter particles to induce detectable recoil in the target of Direct Detection experiments. We derive constraints on scattering cross section of DM and electron, using {\xenon} and {\sc{Super-Kamiokande}} data. Vector and scalar mediators are considered, in the heavy and light regimes. We discuss the importance of including energy dependent cross sections (due to specific Lorentz structure of the vertex) in our analysis, and show that the bounds can be significantly different than the results obtained assuming constant energy-independent cross-section, often assumed in the literature for simplicity. Our bounds are also compared with other astrophysical and cosmological constraints.
\end{abstract}

\maketitle
\section{Introduction}

One of the strongest indicators of Physics Beyond the Standard Model (BSM) is Dark Matter (DM). Its existence can be inferred from diverse observations like galaxy rotation curves, cosmic microwave background radiation (CMBR) and gravitational lensing \cite{Bauer:2017qwy, Bertone:2004pz, Lisanti:2016jxe}. Expectedly, massive experimental and observational efforts have been undertaken to understand its composition and interactions. Moreover, details of structure formation constrain the type of DM and we know that it is only cold dark matter (CDM) which fits all the evidence. However, these observations remain mute about the exact composition of DM and the interactions it has with itself and SM particles besides gravitation. 

Experiments aimed at investigating particle nature of DM are divided into two categories - indirect detection and direct detection (DD) experiments. Indirect detection experiments~\cite{Conrad:2014tla} focus on the study of signatures of the creation or annihilation of DM. Annihilation or decay of DM might produce excess photons in a certain mass window over the SM background, from which the mass of the DM can be inferred. The obvious challenge in this methodology is the very low signal production, which can be difficult to distinguish over the SM  background, not to mention the difficulty
in modelling the SM photon background in the first place. The basic idea of DD experiments is that DM particles impinge on a detector and transfer a part of their kinetic energy to the target. The rate of such scattering events in a certain recoil energy bin yields DM interaction cross section bounds. Despite intense efforts on DD experiments all around the globe, the search for DM has been fruitless. Some experiments have seen tantalising hints~\cite{DAMA:2008jlt,XENON:2020rca}, but nothing definitive has come of those~\cite{XENON:2022mpc}. 

The average velocity of DM particles in the solar neighbourhood is $v/c \sim 10^{-3}$  which limits the energy to be deposited in a detector. Therefore, scattering in the direct detection is assumed to be non-relativistic (NR). With detectors like {\sc{ Xenon1T}} that has a minimum electronic recoil energy threshold of $\sim \mathcal{O}(1~{\rm keV})$, the smallest accessible DM mass ($m_\chi$) is $m_\chi \sim \mathcal{O}(1~{\rm MeV})$. For {\sc{Super-Kamiokande}} ({\sc{Super-K}}), which has a minimum recoil energy threshold of $\sim \mathcal{O}(1~{\text{MeV}})$, the smallest accessible DM mass is $\sim \mathcal{O}(1~{\text{GeV}})$ \footnote{An exception to this occurs in fermionic DM absorption models, for which {\sc{Xenon-1T}} can probe DM masses down to $\sim \mathcal{O}({10\rm{~keV}})$ and {\superk} can probe masses of $\sim \mathcal{O}({1\rm{~MeV}})$~\cite{Dror:2020czw, Dror:2019onn,Dror:2019dib}}. These detectors cannot access lighter DM particles in this scenario. 

However, as DM particles interact with cosmic rays (CR), it is inevitable that some DM particles will 
be boosted due to scattering by energetic CR particles~ \cite{Bringmann:2018cvk,Cappiello:2018hsu,Cappiello:2019qsw,Bringmann:2018cvk,Ema:2018bih,Cappiello:2018hsu,Cappiello:2019qsw,Dent:2020syp,Jho:2020sku,Bramante:2021dyx,Farzan:2014gza,Arguelles:2017atb,Yin:2018yjn,Jho:2021rmn,Das:2021lcr,PhysRevD.105.103029}. In this study we focus on boosting of DM particles by CR electrons only; for boosting of DM by CR nucleons and neutrinos, refer to \cite{Bringmann:2018cvk,Dent:2019krz,Cappiello:2019qsw,Elor:2021swj, Chauhan:2021fzu,PhysRevD.105.103029}. Since boosted particles can carry large amounts of 
kinetic energy, even very light DM particles can deposit a recoil energy $E_R>E_c$ in a 
detector, where $E_c$ is the lower detector threshold. Thus, direct detection detectors as well as 
neutrino experiments can become sensitive to very low mass DM. However, the sensitivity at lower 
DM masses is achieved at larger cross sections because the upscattered subcomponent flux is 
substantially lower than the galactic DM population. Note that CRe is one of sources of boosting DM particles 
among others, such as blazars~\cite{Wang:2021jic,Granelli:2022ysi}, helium nuclei~\cite{Bringmann:2018cvk},
Diffuse Supernova Neutrino Background (DSNB)~\cite{Farzan:2014gza,Arguelles:2017atb,Yin:2018yjn,Jho:2021rmn,Das:2021lcr} and non-galactic contributions to DM flux~\cite{Herrera:2021puj}. 

In most of the existing literature, DM interaction cross-sections have largely been taken to be 
independent of the DM energy. This is a good approximation when i) the DM is non-relativistic and 
ii) the DM mediator is heavy. These assumptions will not hold when : i) DM becomes relativistic 
upon getting upscattered by energetic particles, and ii) mediator is light. The DM boost phenomena 
introduces non-trivial energy dependences for both heavy and light dark mediator \footnote{The mediator is charged under both the SM electroweak group, as well as the DM gauge group, allowing it to couple SM electrons to the DM particles}. The exact energy dependence is 
operator dependent. The importance of energy-dependent scattering has been recently highlighted 
in a few works ~\cite{Dent:2019krz,Cao:2020bwd,Dent:2020syp,Ema:2020ulo,Xia:2022tid}, where 
it was found that the resulting limits are orders of magnitude different than those derived under 
the assumption of a constant cross section.

In this paper we consider the direct detection of DM particles, boosted by Cosmic Ray electron 
(CRe) via recoil of electrons in detectors in specific models of fermionic DM interactions. The 
remainder of this work is organized as follows. In Sec. II we discuss how to obtain the DM flux and 
the event rate from an upscattered DM. Specifically, we show the effects of energy dependence of 
the cross-section on the flux, as induced by the Lorentz structure of the operator, and compare it 
to the boosted constant cross-section case. In Sec. III we consider specific operators. Explicit 
connections of these operators to well-motivated models of DM are also drawn. Sec. IV provides 
the results in the cross-section mass plane from 
{\sc{Super-K}} and {\xenon}, along with a discussion on cosmological constraints from BBN 
as well as collider constraints. We summarise and conclude in Sec.V.

\section{Boosted Dark Matter Flux and Event Rate} 
The DM particles contained in the DM halo within the Milky Way galaxy follow a curtailed Maxwell-
Boltzmann velocity distribution, with the average velocity at about $v \sim 10^{-3}$ (with 
$c = 1$). It is inevitable that the energetic Cosmic Ray electrons will interact with non-relativistic 
DM particles and may provide them a large boost to velocities $v \gg 10^{-3}$. CR electron flux 
($F(T_e)$) can be described by certain parameterization of the local interstellar spectrum~\cite{Boschini:2018zdv} given as
%%%%
\begin{widetext}
\begin{align}
  F(T_e) = \begin{cases}
     \mbox{\Large\( \frac{1.799 \times 10^{44}~ T_e^{-12.061}}{1+ 2.762 \times 10^{36}~ T_e^{-9.269} + 3.853 \times 10^{40}~ T_e^{-10.697}} \)} & \text{if $T_e < 6880$ MeV} \\
   \mbox{\tiny\( \)} & \mbox{\tiny\( \)} \\
      3.259 \times 10^{10}~T_e^{-3.505} + 3.204 \times 10^{5}~T_e^{-2.620} & \text{if $T_e \geqslant 6880$ MeV}
    \end{cases}
\label{Eq:CRe_flux_parameterization}
\end{align}
\end{widetext}
%%%% 
where the unit of $F(T_e)$ is given in $\rm{\left(m^2~s~sr~MeV \right)^{-1}}$ and the kinetic energy of the CR electrons ($T_e$) is in MeV. The above fit is consistent with Fermi-LAT~\cite{Fermi-LAT:2011baq,Fermi-LAT:2009yfs,Fermi-LAT:2010fit,Fermi-LAT:2017bpc}, AMS-02~\cite{AMS:2014gdf}, PAMELA~\cite{PAMELA:2011bbe,CALET:2017uxd}, and Voyager~\cite{Cummings:2016pdr,Stone:2013} local interstellar spectrum data, to within an accuracy of $5$ \% .

For a CR electron (CRe) hitting a DM particle, we have
\begin{eqnarray}
T^{\rm max}_\chi  &=& \frac{T_{e}^{2}+2 m_{e} T_{e}}{T_{e}+\left(m_{e}+m_{\chi}\right)^{2} /\left(2 m_{\chi}\right)}\label{eqn:Tchimax} \\
T_\chi &=& T_\chi^{\rm max} \frac{1-\cos \theta}{2}
\end{eqnarray}
where $T_{\chi} (T_e)$ is the kinetic energy of the DM particle (CRe), $m_{\chi} (m_e)$ is the mass of the DM particle (CRe) and $\theta$ is the scattering angle in the centre of momentum frame. 

The differential flux of the Boosted DM (BDM) is then given by
\begin{equation}
\left(\frac{d \Phi_{\chi}}{d T_{\chi}}\right)_{e}=D_{\rm e f f} \times \frac{\rho_{\chi}^{\text {local }}}{m_{\chi}} \int_{T_{e}^{m i n}(T_{\chi})}^{\infty} d T_{e} \frac{d \Phi_{e}}{d T_{e}} \frac{d \sigma_{\chi e}}{d T_{\chi}}
\end{equation}
where $\Phi_\chi (\Phi_e)$ is the DM (CRe) flux, $\rho_\chi^{\rm local}$ is the local DM density, $\sigma_{\chi e}$ is the DM-CRe interaction cross-section, $D_{\rm eff}$ {\footnote {We need to consider all possible line segments along the line of sight, along which the DM particles are boosted after the interaction with CRe. $D_{\text{eff}}$ is the effective distance out to which all CRe have to be taken into account}} is the line-of-sight effective distance (taken to be $1~\text{kpc}$), and $T_e ^{\text{min}}$ is the minimum kinetic energy CRe must possess to boost the DM particle to energy $T_{\chi}$, given by :

\begin{eqnarray}
\hspace {-3mm} T_{e}^{\text{min}}=\left( \frac{T_{\chi}}{2}-m_{e} \right) \left[1\pm \sqrt{1+\frac{2T_{\chi}}{m_{\chi}} \frac{(m_e+m_{\chi})^2 }{(2m_{e}-T_{\chi})^2}} ~\right]
\end{eqnarray}
with $+$ and $-$ applicable for $T_{\chi}>2m_e$ and $T_{\chi}<2m_e$ respectively.

 Of course, 
\begin{equation}
\frac{d \sigma_{\chi e}}{d T_{\chi}}=\frac{|\mathcal{M}|^{2}}{16 \pi s_{\text{CR}}} \frac{1}{T_{\chi}^{\max }}
\label{eqn:diffxsec}
\end{equation}
where $\mathcal{M}$ is the interaction matrix element and $s_{\text{CR}}$ is the centre of momentum energy for the CRe-DM collision, given by :
\begin{eqnarray}
s_{\text{CR}}=\left(m_\chi +m_e\right)^2 + 2m_\chi T_{\text{e}}
\label{eqn:sCR}
\end{eqnarray} 

Under the energy independent approximation for the cross section, the differential cross section would simply be :
\begin{eqnarray}
\frac{d\sigma_{\chi e}}{dT_{\chi}}=\frac{ \bar{\sigma}_{e \chi}}{T_{\chi} ^{\text{max}}}
\end{eqnarray}

We define the following quantities : 
\begin{eqnarray}
\mathbb{M}^2 &=&\frac{16 g_{e}^{2} g_{\chi}^{2} m_{e}^{2} m_{\chi}^{2}}{\left(q_{\mathrm{ref}}^{2}-m_{i}^{2}\right)^{2}}\\
\bar{\sigma}_{e \chi} &=& \frac{\mu_{\chi e}^2}{16 \pi m_e^2 m_\chi^2} \mathbb{M}^2
\end{eqnarray}
where $q_{\rm ref} = \alpha m_e$ is the reference momentum transferred. Here, $g_\chi$ ($g_e$) is the coupling constant
of the dark mediator to the DM particle  (electron), $m_i$ is the mass of the dark mediator ($i=A',\phi$ for vector, scalar mediator) and $\mu_{e\chi}$ is the reduced mass of the DM-electron system. 

The differential cross-section is given by
\begin{equation}
\frac{d \sigma_{\chi e}}{d E_{R}} = \frac{|\mathcal{M}|^{2}}{16 \pi s_\chi} \frac{1}{E_R^{\max }}
\label{eqn:diffxsecER}
\end{equation}
where $s_\chi$ is centre of momentum energy for the DM-target electron collision which can be obtained from Eqn.~\eqref{eqn:sCR} under the substitution : $m_\chi\leftrightarrow m_e$ and $T_{\text{e}}\rightarrow T_\chi$ . $E_R ^{\text{max}}$ is the maximum possible recoil in the detector, that can be imparted by a DM particle with kinetic energy $T_\chi$, and can be obtained from Eqn.~\eqref{eqn:Tchimax} with the appropriate substitutions mentioned before.

We can now define a form factor 
\begin{equation}
F_{\rm DM}^2 (q^2) = |\mathcal{M}|^{2} / \mathbb{M}^2
\label{eqn:fdmsq}
\end{equation}
This factor contains the energy dependence arising in the differential cross section $d \sigma_{\chi e}/ d T_{\chi}$ due to CRe boosting the DM particles and the Lorentz structure of the interaction. The explicit form of $F_{\rm DM}$ depends on the model of DM and mediator considered.

A similar form factor, $F_{\rm rec}$, contains energy dependence in the differential cross section $d \sigma_{\chi e}/ d E_{R}$ arising due to interaction of relativistic DM particles with the electrons in the detector, and can be obtained from the form factor $F_{\rm DM}$ of Eqn.~\ref{eqn:fdmsq} by making the substitutions : $m_e \leftrightarrow m_\chi$, $T_\chi \rightarrow E_R$ and $T_e \rightarrow T_\chi$.

Hence the differential cross sections, $d \sigma_{\chi e}/ d T_{\chi}$ and $d \sigma_{\chi e}/ d E_{R}$, relevant in the DM-CRe scattering and DM scattering at the detector end respectively, are given by :

\begin{eqnarray}
\frac{d\sigma_{\chi e}}{dT_{\chi}}= \bar{\sigma}_{e \chi} \frac{m_e ^2 m_{\chi} ^2}{\mu _{e \chi} ^2} \frac{ F_\text{DM} ^2 (q^2)}{s_{\text{CR}} T_{\chi} ^{\text{max}}} 
\label{eqn:dsig/dT}
\end{eqnarray}

and,

\begin{eqnarray}
\frac{d\sigma_{\chi e}}{dE_{R}}= \bar{\sigma}_{e \chi} \frac{m_e ^2 m_{\chi} ^2}{\mu _{e \chi} ^2} \frac{ F_\text{rec} ^2 (q^2)}{s_\chi E_{R} ^{\text{max}}} 
\label{eqn:dsig/dER}
\end{eqnarray}

The differential recoil rate of electrons in {\sc{Super-K}} can be calculated to be
\begin{eqnarray}
\frac{d R}{d E_{R}} &=& \aleph \int_{T_{\chi}^{\min }\left(E_{R}\right)}^{\infty} d T_{\chi}\left(\frac{d \Phi_{\chi}}{d T_{\chi}}\right)_{e} \frac{d \sigma_{\chi e}}{d E_{R}}
\end{eqnarray}
where the factor $\aleph$ takes into account the number density of the target electrons in the detector, $E_R$ is the recoil energy and $T_{\chi}^{\text{min}}$ is the minimum DM energy required to produce a recoil of $E_R$ in the detector, given by 
\begin{eqnarray}
T_{\chi}^{\text{min}}=\left( \frac{E_{R}}{2}-m_{\chi} \right) \left[1\pm \sqrt{1+\frac{2E_{R}}{m_{e}} \frac{(m_e+m_{\chi})^2 }{(2m_{\chi}-E_R)^2}} ~\right] \nonumber \\
\end{eqnarray}
with $+$ and $-$ applicable for $E_{R}>2m_{\chi}$ and $E_{R}<2m_{\chi}$ respectively.

The detection mechanism for {\xenon} detector consists of an ionisation process. In the {\xenon} detector, an incident DM particle can ionize an electron in the $(n,l)$ shell of a Xenon atom ($A$). The rate of the ionization process $\chi + A \rightarrow \chi + A^+ + e^-$ is given by

\begin{eqnarray}
\frac{dR_{\text{ion}}}{d \ln E_R} = \tilde{\aleph} \phi_{\text{halo}} \sum_{nl} \frac{d \langle \sigma_{\text{ion}}^{nl} v \rangle}{d \ln E_R}
\end{eqnarray}

where $\tilde{\aleph}$ is the number of target atoms in the detector, $\phi_{\text{halo}}=n_\chi \bar{v}_\chi$ is the background galactic DM halo flux, and $\frac{d \langle \sigma_{\text{ion}}^{nl} v \rangle}{d \ln E_R}$ is the velocity-averaged differential cross section, given by \cite{Essig:2011nj,Essig:2015cda} :

\begin{eqnarray}
\frac{d \langle \sigma_{\text{ion}}^{nl} v \rangle}{d \ln E_R} = \frac{\bar{\sigma}_{e \chi}}{8 \mu^2_{\chi e}} \int |F_{\text{rec}}(q)|^2 |f_{\text{ion}} ^{nl} (k',q)|^2 \eta(E_\chi^{\text{min}}) q dq , 
\nonumber \\
\label{eq:sigmav_ion}
\end{eqnarray}

Here $F_{\text{rec}}$ is a form factor defined and discussed below (See Eqns.~\ref{eqn:fdmsq},~\ref{eqn:dsig/dER}), $f_{\text{ion}} ^{nl} (k',q)$ is the ionization form factor and $q$ is the momentum transferred  (See Appendix~\ref{appendix:A} for more details). The mean inverse speed function $\eta(E_\chi^{\text{min}})$ is given by \cite{An:2017ojc} 
\begin{eqnarray}
\eta(E_\chi ^\text{min}) = \int_{E_\chi ^{\text{min}}} dE_\chi \phi_{\text{halo}}^{-1} \frac{m_\chi ^2}{p E_\chi} \frac{d\phi_\chi}{dT_\chi}
\label{eq:eta}
\end{eqnarray}

where $E_\chi ^\text{min}$ refers to the minimum energy that a DM particle must possess to elicit the detector recoil $E_R$. Note that $E_\chi ^\text{min} = T_\chi ^{\text{min}} +m_\chi$. Also note that this convolution need not be done for {\superk}.

\begin{figure}[!h]
\begin{subfigure}{\linewidth}
\hspace*{-6 mm}
\includegraphics[scale=0.6]{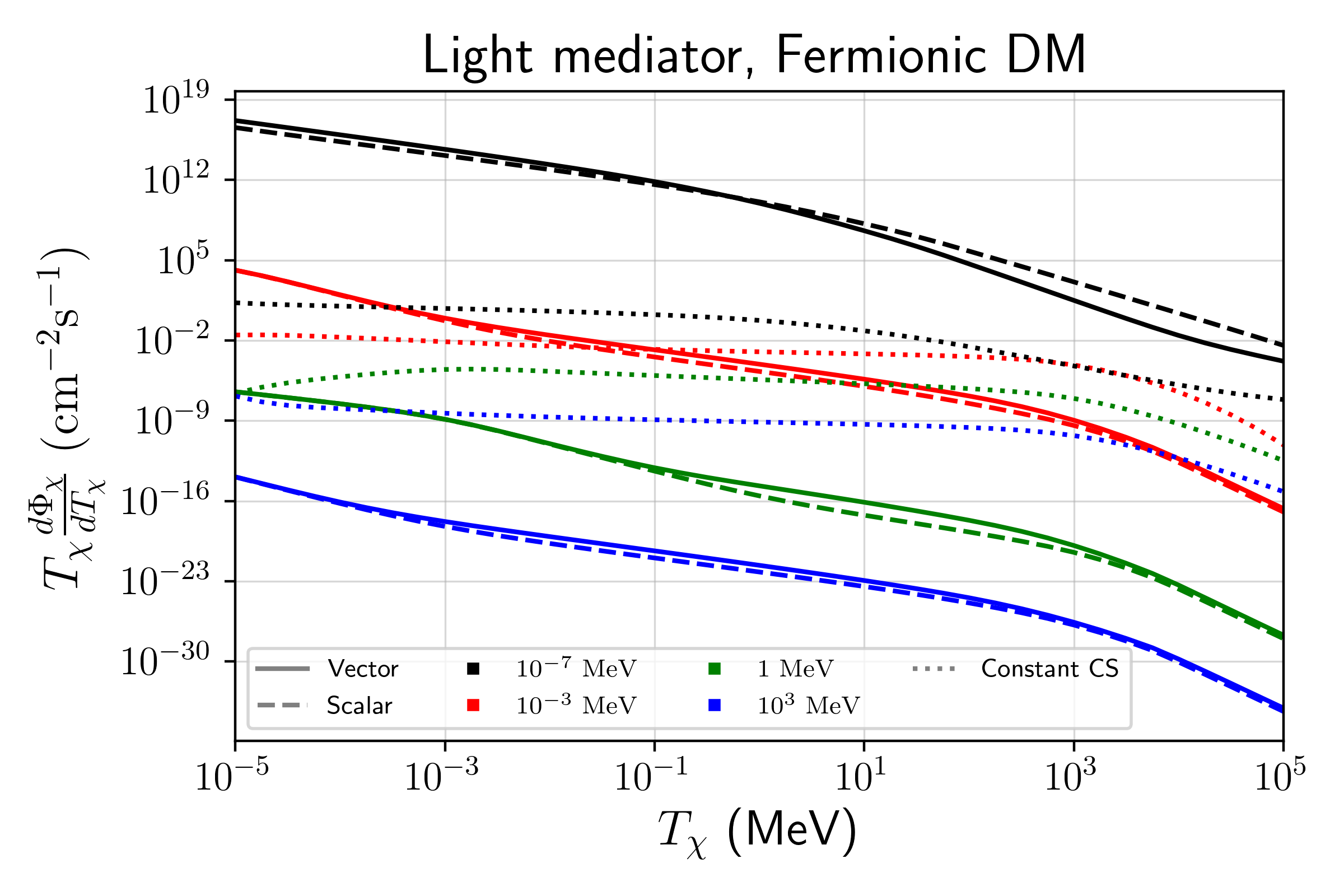}
\caption{Light mediator}
\label{fig:fluxlight}
\end{subfigure}
\begin{subfigure}{\linewidth}
\hspace*{-6 mm}
\includegraphics[scale=0.6]{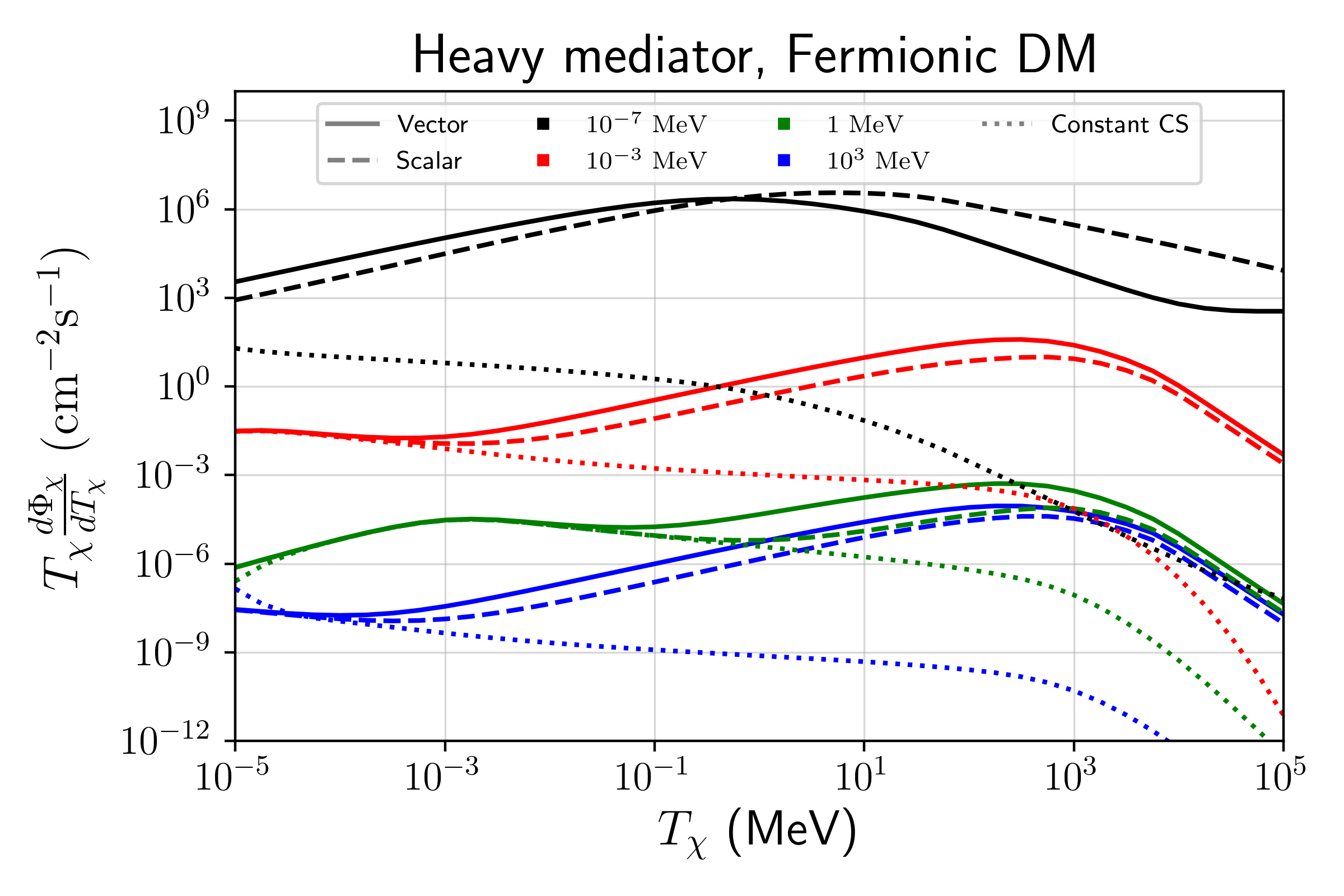}
\caption{Heavy mediator}
\label{fig:fluxheavy}
\end{subfigure}
\caption{Plots showing the effect of Lorentz structure of the operators compared with the constant cross-section ($\bar{\sigma}_{e\chi} = 10^{-30} {\rm cm^2}$) case, on boosted DM flux. For each case, we plot lines for four DM masses, $m_\chi = 10^{-7}, 10^{-3}, 1, 10^3 {\rm \ MeV}$. For the light mediator (Fig.~\ref{fig:fluxlight}), the modified flux is raised above the constant cross-section case for very light DM masses, while it falls below that for higher masses. However, for the heavy mediator (Fig.~\ref{fig:fluxheavy}) case, the modified flux is higher than the constant cross-section case for all DM masses.}
\label{fig:DM_boost}
\end{figure}

The effect of the energy dependence on DM flux can be understood from Fig.~\ref{fig:DM_boost}. The minimum energy the DM particles must possess, in order to impart a detectable recoil to the target electrons, sets the lower limit ($T_{\chi}^{\text{min}}$) of the relevant DM energy range. Very high DM energies ($T_{\chi} >10^3~\text{MeV}$) are not relevant, since the differential flux ($d\Phi_\chi/dT_\chi$) falls off at high DM energies. 

For heavy mediators (vector and scalar), the boost is more effective in increasing the flux at high DM energies when energy dependence of cross section is taken care of. This is applicable for all DM masses, hence it is expected that including energy dependence for heavy mediator will improve the bounds as compared to the constant cross section scenario. For light mediators (vector and scalar), the energy dependent boost is less effective than constant cross section scenario for higher DM masses. This allows us to predict that the light mediator bounds will be stronger than the energy independent bounds for lighter DM, but the same will become weaker for heavier DM. Also, since in the $T_{\chi}$ regime relevant to us, the vector mediator boosted DM flux is greater than the scalar case, we can expect exclusion bounds to be stronger for the former. Finally, since the flux falls for heavier DM, we expect exclusion bounds to be stronger for lighter DM. We find, in Section~IV, that the exclusion bounds we obtain follow these trends.   

\section{Simplified Model and Effective Operators}
Without referring to an underlying model, we consider a fermionic DM particle $\chi$ of mass $m_\chi$, which couples to electrons only. This type of scenario can arise in several
leptophilic models of particle DM~\cite{Pospelov:2007mp,Batell:2009di,Chu:2011be,Alves:2013tqa,Izaguirre:2013uxa,Izaguirre:2015yja,Krnjaic:2015mbs,Izaguirre:2017bqb,Harigaya:2019shz,Bernreuther:2020koj}. For concreteness, we assume this interaction is mediated by a scalar ($\phi$) or a vector mediator ($B_\mu$). 
\begin{eqnarray}
 \mathcal{L} &=& g_{\chi \phi} \phi \bar{\chi}{\chi} + g_{e \phi} \phi \bar{e} e \quad \,\text{or}\\ 
  &=& g_{\chi A^\prime} A^\prime_\mu \bar{\chi}\gamma^\mu{\chi} + g_{e A^\prime} A^{\prime}_\mu \bar{e}\gamma^\mu e
\end{eqnarray}
Depending on the type of operator, we expect  the differential rates to change. In this section, we inspect the effect of the Lorentz structure on $F_{\rm DM}^2(q^2)$ and on the differential rate. 

\subsection{Scalar Mediator}
Considering a scalar mediator (denoted as $\phi$), one can calculate $F_{\rm DM}^2$ for the interaction between CRe and non-relativistic DM, using Eqn.~\ref{eqn:fdmsq} to obtain
\begin{equation}
F_{\mathrm{DM}}^{2}(q)=\frac{\left(q_{\mathrm{ref}}^{2}-m_{\phi}^{2}\right)^{2}}{\left(q^{2}-m_{\phi}^{2}\right)^{2}} \frac{\left(2 m_{\chi}+T_{\chi}\right)\left(2 m_{e}^{2}+m_{\chi} T_{\chi}\right)}{4 m_{\chi} m_{e}^{2}}
\label{eqn:ff-scalar}
\end{equation}

The differential cross section ($d\sigma/dT_{\chi}$) w.r.t. the DM energy ($T_{\chi}$), is :
\begin{eqnarray}
\frac{d\sigma_{\chi e}}{dT_{\chi}} &=& \bar{\sigma}_{e \chi}  \frac{(q_{\mathrm{ref}}^{2}-m_{\phi}^{2})^{2}}{(q^{2}-m_{\phi}^{2})^{2}} 
\left\{ \frac{m_{\chi}}{4 \mu _{e \chi} ^2}\right.
\nonumber \\
&&\left.  \frac{\left(2 m_{\chi}+T_{\chi}\right)\left(2 m_{e}^{2}+m_{\chi} T_{\chi}\right)}{ s_{\text{CR}} T_{\chi} ^{\text{max}}} \right\}
\label{eqn:sigma-scalar}
\end{eqnarray}

The form factor $F_{\text{rec}}$ and the differential cross-section w.r.t. the recoil energy of the detector ($d\sigma_ {\chi e}/dE_R$) are obtained from Eqn.~\eqref{eqn:ff-scalar} and Eqn.~\eqref{eqn:sigma-scalar} by performing the substitutions prescribed in the previous section, viz. $m_e \leftrightarrow m_\chi$, $T_\chi \rightarrow E_R$, $T_e \rightarrow T_\chi$, $s_{\text{CR}} \rightarrow s_{\chi}$ .

\subsection{Vector Mediator}
Using a similar treatment for the vector mediator (denoted by $A'$), we find that
\begin{eqnarray}
F_{\mathrm{DM}}^{2}(q^2)&=&\frac{\left(q_{\mathrm{ref}}^{2}-m_{A'}^{2}\right)^{2}}{\left(q^{2}-m_{A'}^{2}\right)^{2}} \frac{1}{2 m_{\chi} m_{e}^{2}} \left( 2 m_{\chi}\left(m_{e}+T_{e}\right)^{2}-\right. \nn \\ 
&& \left. T_{\chi}\left\{\left(m_{e}+m_{\chi}\right)^{2}+2 m_{\chi} T_{e}\right\}+m_{\chi} T_{\chi}^{2}\right)
\end{eqnarray}
and,
\begin{eqnarray}
\frac{d\sigma_{\chi e}}{dT_{\chi}} &=& \bar{\sigma}_{e \chi} \frac{\left(q_{\mathrm{ref}}^{2}-m_{A'}^{2}\right)^{2}}{\left(q^{2}-m_{A'}^{2}\right)^{2}} \frac{m_{\chi}}{2 \mu _{e \chi} ^2 s_{\text{CR}} T_{\chi} ^{\text{max}}}\left\{ 2 m_{\chi}(m_{e}+ T_{e})^{2}  \right.
\nonumber \\
&&\left. - T_{\chi} \{ (m_{e}+m_{\chi})^{2}+2 m_{\chi} T_{e}\} + m_{\chi} T_{\chi}^{2}\right\}
\label{eqn:sigma-vector}
\end{eqnarray}

\section{Results}
In this section, we performed a $\chi^2$ analysis to obtain novel limits using {\xenon} (a low energy threshold recoil experiment) and {\sc{Super-K}} (a high energy threshold recoil experiment) data. 

The exclusion region is obtained using the following definitions for $\chi^2$:
 \begin{eqnarray}
 \chi^2 &=& \sum_{i} \frac{(O_i - E_i)^2}{\left(\sigma_i\right)^2 _{\rm{data}}} \\
 \Delta \chi^2 &=& \chi^2(\text{BDM}+\text{B}_0)-\chi^2(\text{B}_0~\rm{only})
 \end{eqnarray}
 
where, $O_i$ are the observed number of events, $E_i$ are the expected number of events and $(\sigma_i)_{\text{data}}$ is uncertainity in the measured data, for the $i^{\rm th}$ recoil energy bin. For the $(\text{BDM}+\text{B}_0)$ case, to calculate the $E_i$ values, we sum the BDM signal and the background $\text{B}_0$ for each energy bin. Clearly, if the BDM contribution explains experimental data, $\Delta \chi^2$ must be less than $0$ corresponding with a better fit. 

The {\sc Xenon1T} collaboration had reported a $3.5\sigma$ excess of events in the electron recoil range of $1~\rm{keV} < E_R < 7~\rm{keV}$ \cite{XENON:2020rca}. However, a recent
dataset from the {\xenon} experiment \cite{XENON:2022mpc}, aimed at verifying the
aforementioned excess, shows that no such excess exists. We use the data from this experiment for our analysis. To derive the exclusion limit with the 95\% confidence, we demand $\Delta \chi^2 > 40.1$ that corresponds to 27 degrees of freedom.

For {\superk}, we use the SK-I data which was taken for total 1497 days of live-time~\cite{Super-Kamiokande:2011lwo}. The detector originally looked for the Diffuse Supernovae Background events via inverse beta decay $\bar{\nu_e} + p \to n + e^+ $. In the present work, we assume that the observed events are consistent with the background and hence the signal due to DM should be consistent with the data within the uncertainity. Since an estimate of the background is not found in the literature for SK-I data, we take $\chi^2(B_0~\rm{only})|_{\rm SK} = 0$. The excluded region satisfies $\Delta \chi^2 > 26.3$ which corresponds to 95\% exclusion limit for
16 degrees of freedom.
 
Both {\superk} and {\xenon} experiments are located deep underground to reduce background, but this also attenuates the DM flux entering the detector. The attenuation of DM particles happens mainly due to the interaction with electrons in the Earth's surface, significantly altering the DM flux reaching the detector. While a detailed study of the effects of attenuation on boosted DM is beyond the scope of this paper, we have determined the attenuation bound considering a DM particle with $T_\chi=1~\text{GeV}$. This attenuation bound corresponds to the cross section for which the DM particle (with $T_\chi =1~\text{GeV}$)  can impart the threshold recoil energy in the detector. For this, we solve the following equation to calculate the energy $T_r$ lost by the dark matter

\begin{eqnarray}
\label{eq:Tchiz}
\frac{dT_\chi}{dx} = -\sum_T n_T \int_0 ^{T_r ^{\text{max}}} \frac{d\sigma}{dT_r}T_rdT_r
\end{eqnarray}

and estimate $\bar{\sigma}_{e \chi}$ so that kinetic energy of the DM particle at depth $z$, denoted by $T_\chi ^z$, is the detector threshold $E_{\rm th}$, for an initial kinetic energy $T_{\chi,{\rm{in}}}=1~\text{GeV}$. The area bounded by the attenuation bound and the exclusion bound is ruled out by our analysis. Also note that ionisation effects could dominate above $T_{\chi} =$ 1 GeV. Moreover, light DM particles ($m_{\chi} < m_e$) may backscatter into the atmosphere. In this work, though, we limit ourselves to elastic scattering, leaving a more elaborate treatment for future work. Note that the attenuation limits exist only for the heavy mediators. There is no attenuation bound shown for the light mediator scenario with elastic scatterings and the attenuation bound shown for heavy mediator may also vary once the effects mentioned above are taken into account.

\begin{figure}[!h]
\begin{subfigure}{\linewidth}
\hspace*{-7mm}
\includegraphics[scale=0.6]{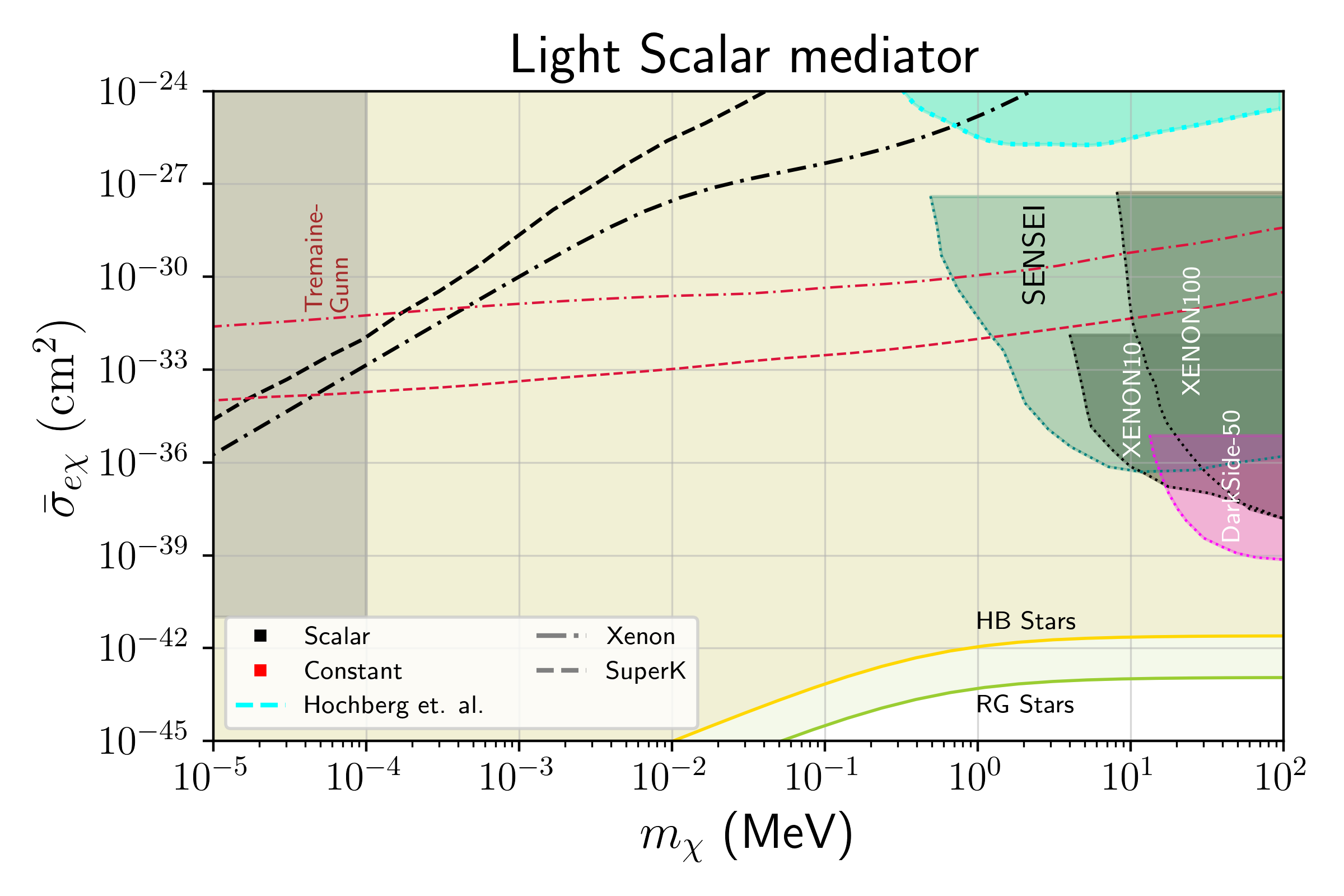}
\caption{Light Scalar mediator}
\label{fig:light_scal_med_bound}
\end{subfigure}
\begin{subfigure}{\linewidth}
\hspace*{-7mm}
\includegraphics[scale=0.6]{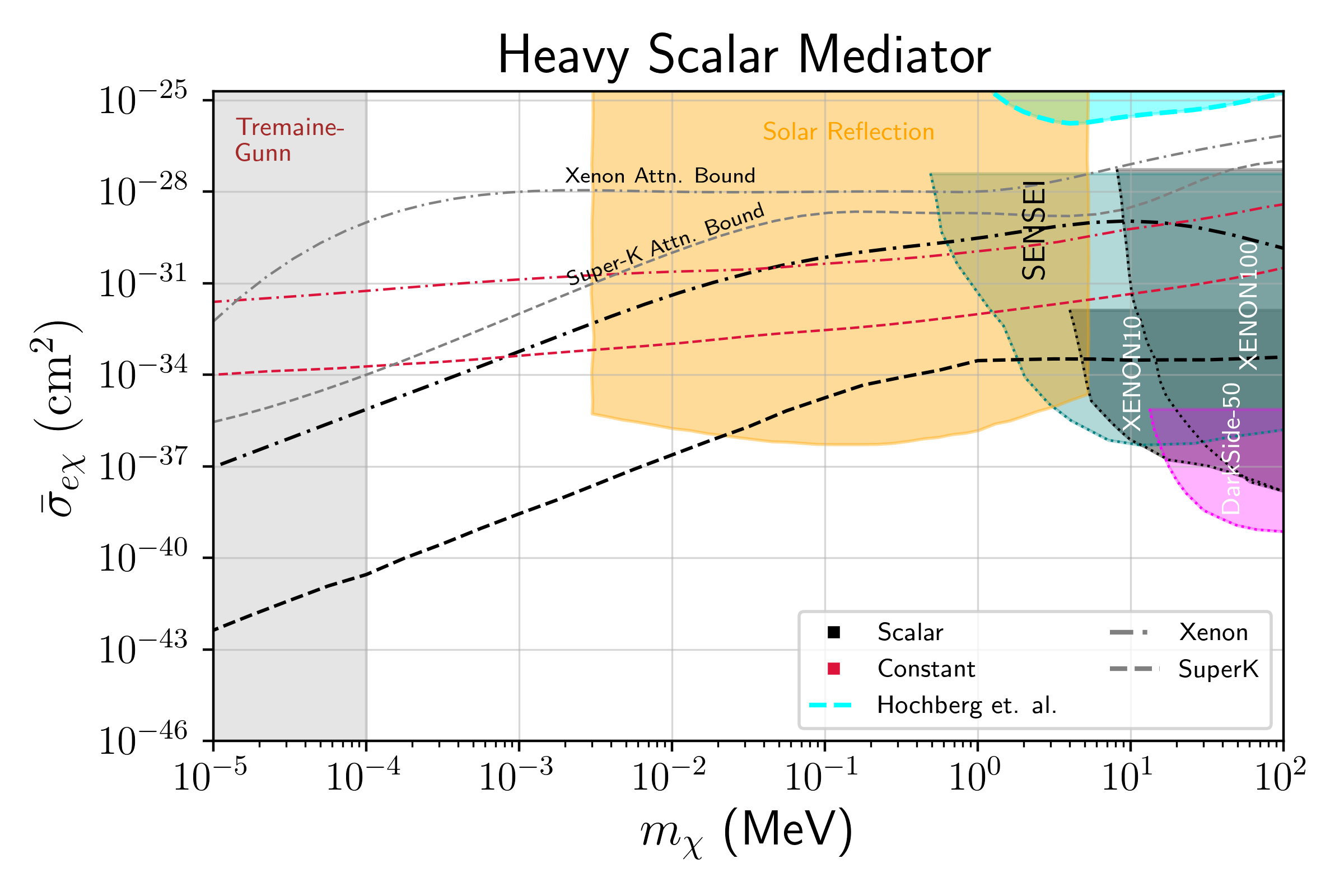}
\caption{Heavy Scalar mediator}
\label{fig:heavy_scal_med_bound}
\end{subfigure}
\caption{Exclusion bounds on the cross-section is shown as a function of the DM mass for the scalar mediator. Exclusion bound for constant cross-section scenario is also plotted (in red). For each of these scenarios, the results are shown for two different experiments - {\xenon} and {\superk}, differentiated by the linestyles used in the plot. The direct detection bounds from {\sc Xenon10,Xenon100, SENSEI}~\cite{Essig:2012yx,Essig:2017kqs,SENSEI:2020dpa} and {\sc DarkSide-50}~\cite{DarkSide-50:2022hin} are also plotted. The grey shaded region represents the region excluded due to the Tremaine-Gunn bound. The bound arising due to DM attenuation is also given for heavy mediator scenario. Note that the region between attenuation bound and exclusion bound is ruled out. Bounds from stellar cooling constraints \cite{Hardy:2016kme} are also shown for light mediator case, while for the 
heavy mediator case, the bound from solar reflection of DM \cite{An:2017ojc,Cao:2020bwd} is shown. }
\label{fig:exclusion_scal}
\end{figure}

\begin{figure}[!h]
\begin{subfigure}{\linewidth}
\hspace*{-7mm}
\includegraphics[scale=0.6]{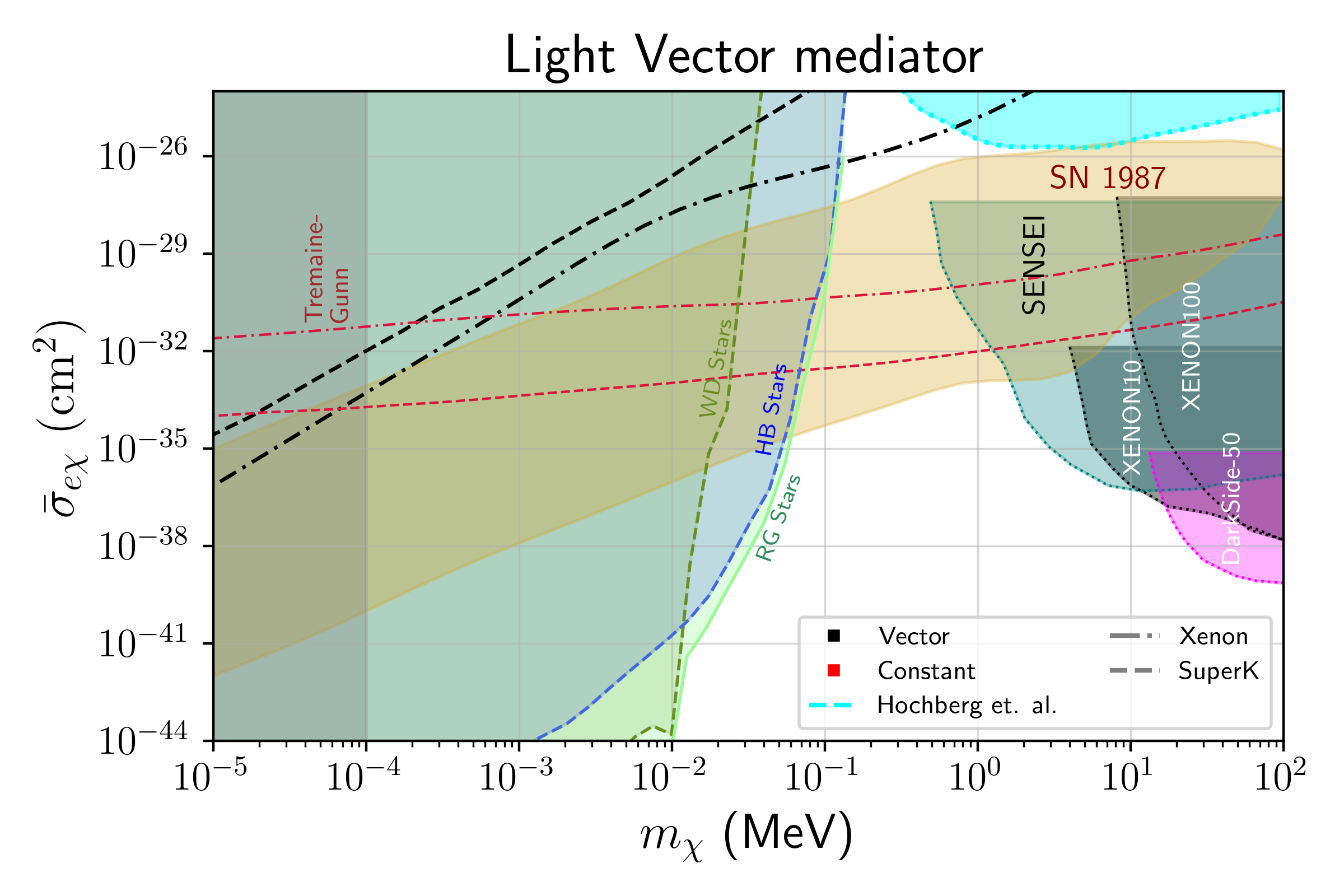}
\caption{Light Vector mediator}
\label{fig:light_vec_med_bound}
\end{subfigure}
\begin{subfigure}{\linewidth}
\hspace*{-7mm}
\includegraphics[scale=0.6]{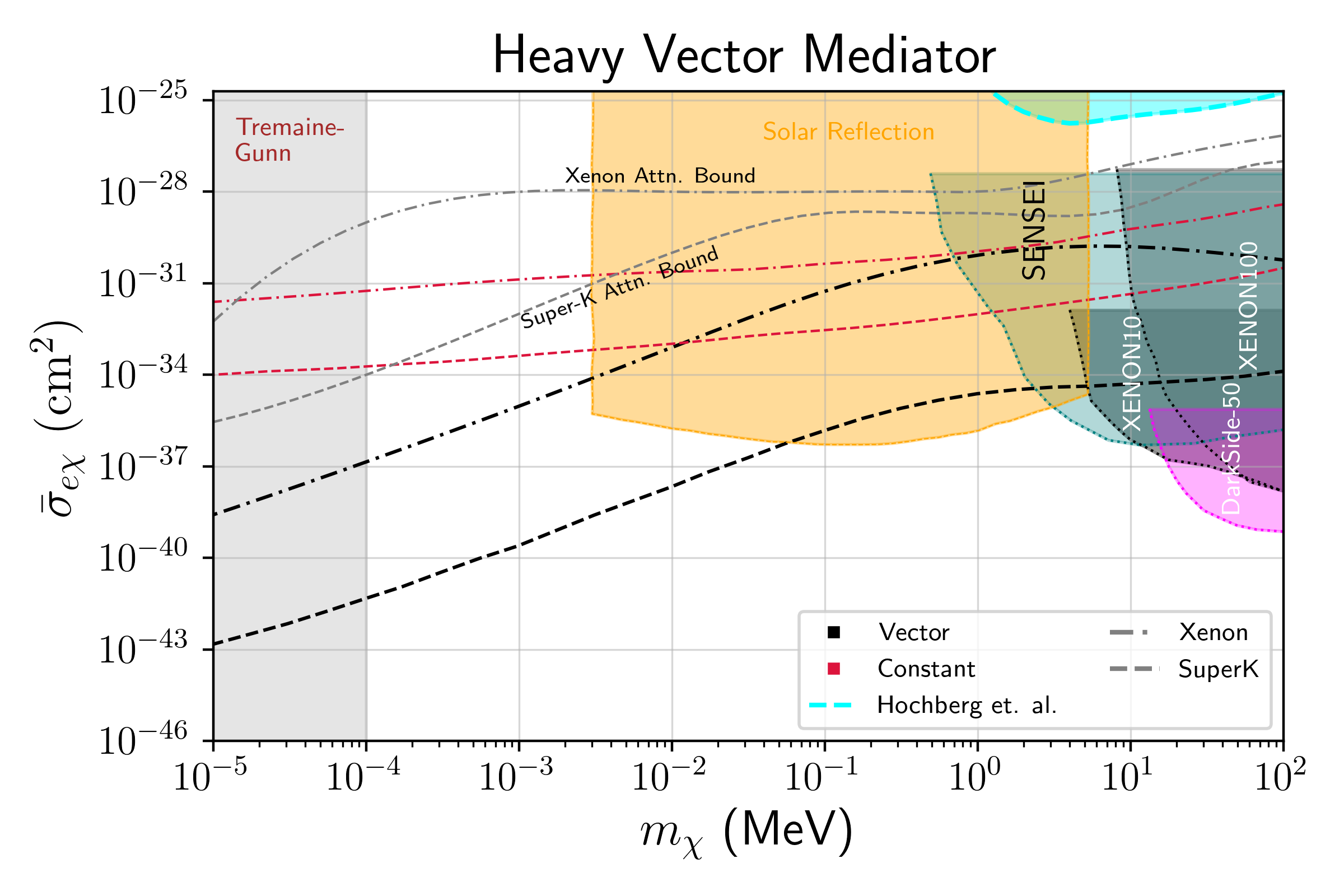}
\caption{Heavy Vector mediator}
\label{fig:heavy_vec_med_bound}
\end{subfigure}
\caption{Exclusion bounds on the cross-section is shown as a function of the DM mass for the vector mediator. Exclusion bound for constant cross-section scenario is also plotted (in red). For each of these scenarios, the results are shown for two different experiments - {\xenon} and {\superk}, differentiated by the linestyles used in the plot. The direct detection bounds from {\sc Xenon10,Xenon100, SENSEI}~\cite{Essig:2012yx,Essig:2017kqs,SENSEI:2020dpa} and {\sc DarkSide-50}~\cite{DarkSide-50:2022hin} are also plotted. The grey shaded region represents the region excluded due to the Tremaine-Gunn bound. The bound arising due to DM attenuation is also given for heavy mediator scenario. Bounds from stellar and supernovae (SN 1987) cooling \cite{Hardy:2016kme,Chang:2018rso} are also shown for light mediator case. Constraint due to solar reflection of DM, relevant for the heavy mediator case \cite{An:2017ojc,Cao:2020bwd}, is shown in amber color. }
%Solar production constraints are obtained by requiring that the dark photon luminosty does not exceed $10 \%$ of the standard solar luminosity \ref{An:2013yua}. }
\label{fig:exclusion_vec}
\end{figure}

The exclusion bounds arising from {\xenon} and {\superk} data are shown in Fig.~\ref{fig:exclusion_scal},~\ref{fig:exclusion_vec} in the heavy and light mediator regime for scalar and vector operators. We find that {\superk} sets the stronger bound for heavy scalar and vector mediators. For light mediators, it is {\xenon} that sets the stronger bound, even though {\superk} has a greater live-time and a larger effective target density $\aleph$. Fermionic DM lighter than $\mathcal{O}(100 \rm{~eV})$ is highly constrained by the Tremaine-Gunn bound \cite{PhysRevLett.42.407, DiPaolo:2017geq, Savchenko:2019qnn,Pal:2019tqq}\footnote{For a possible way to evade this bound, see Ref.~\cite{Davoudiasl:2020uig}}.  
We find that, for the light mediator case, the energy dependent cross section bounds are 
stronger than the constant cross section bounds for keV-scale DM, and weaker for heavier DM. For the 
heavy mediator case, the energy-dependent bound is stronger than the constant cross-section 
case and competitive for heavier DM. This is, as previously discussed in Section II, a consequence 
of BDM flux behaviour, shown in Fig.~\ref{fig:DM_boost}. Ofcourse, the exact value of DM mass at 
which energy independent cross section bounds take over as $m_{\chi}$ is increased cannot be 
predicted by the flux plots alone, since there is $T_{\chi}$ dependence in differential cross section 
relevant at the detector end as well. As discussed in Section~II, the vector mediator case yields 
slightly stronger bounds than the scalar mediator case. We also plot the DM attenuation bound for {\xenon} and {\sc{Super-K}}, for heavy mediator scenario.

Similar results have been obtained for vector mediator in Ref.~\cite{Cao:2020bwd}, but it should be noted that the dataset
used  in Ref.~\cite{Cao:2020bwd} is based on {\sc Xenon100} and {\sc Xenon1T} 's S2-only analysis \cite{XENON:2016jmt,XENON:2019gfn}, while we use the data based on {\xenon} 's S1-S2 analysis and thus the exclusion bounds we provide are slightly different from those obtained in Ref.~\cite{Cao:2020bwd}.

Our bounds for boosted DM can also be compared to bounds obtained for non-relativistic DM using
novel materials with extremely low recoil trigger. A prototype device that can measure single
photons made using Superconducting Nanowires is described in Refs.~\cite{Hochberg:2019cyy, Hochberg:2021yud}.  The best bounds obtained from the device is also shown. The bounds they obtained are competitive with our bounds for DM masses $m_\chi \gtrsim \mathcal{O}(1 \ {\rm MeV})$. At the moment
our bounds are much stronger for lower masses, but proposed devices with materials like NbN and Al might give
better exclusions in the near future. 

We have included constraints arising from astrophysical sources like Red Giant and Horizontal Branch stars~\cite{Hardy:2016kme} for light mediators. For light scalar mediators, stellar cooling bounds are so severe that they rule out the whole region constrained in this work. In case of a vector mediator, bounds are mild for ultra-light mediator due to in-medium effects~\cite{Vogel:2013raa,Knapen:2017xzo}. Bounds from solar reflection of DM \cite{An:2017ojc,Cao:2020bwd} are important in the 
heavy mediator case. The  cosmological constraints from Big Bang Nucleosynthesis (BBN) rule out thermal DM of $m_\chi \lesssim 10{\ \rm MeV}$ stringently~\cite{Knapen:2017xzo,Ghosh:2020vti}. Similarly, the CMB observations constrain DM annihilating to an $e^-e^+$ pair severely~\cite{Planck:2018vyg}. However, BBN bounds are relaxed in models where DM couples to both neutrinoes and electrons~\cite{Escudero:2018mvt}. Also, if there is an elaborate dark sector associated in these models so that DM mostly annihilate to other dark sector particles, BBN and CMB constraints can be relaxed even further~\cite{Choudhury:2020xui}.  For the heavy mediator case,  some of the proposed or approved future experimental facilites and detection strategies, discussed in Ref.~\cite{Batell:2022dpx}, have great potential to explore the parameter space probed by {\xenon} and {\sc {Super-K}} shown in Fig.~\ref{fig:exclusion_scal} and Fig.~\ref{fig:exclusion_vec}.

%\sout{we find that the {\sc BaBar} searches for dark photons and dark scalars via $e^-e^+\to\gamma A'$~\cite{Lees:2014xha,Lees_2017} constrain $g_{e\,i}\sim\,10^{-3}$. Note that, in general, bounds from various beam-dump experiments like E141~\cite{Riordan:1987aw}, E137~\cite{Bjorken:1988as,Batell:2014mga}, Orsay linac~\cite{Davier:1989wz}, NA64~\cite{Banerjee:2016tad,Banerjee:2017hhz} could be important but they are irrelevant for the chosen benchmark points. However,}

\section{Summary \& Outlook}

Dark Matter (DM) poses a unique challenge in physics at the moment. On one hand, a lot of
cosmological evidence points to its existence, but, on the other hand, its particle nature is
completely unknown. Detection of DM has primarily relied on large terrestrial experiments 
with a lot of targets for a DM particle from the Milky Way galactic halo to impinge on. These
direct detection experiments can then measure the recoil of the target and thus measure both the kinetic energy and mass of the DM particle. 

The challenge to this strategy comes from the fact that DM in our galactic halo is
non-relativistic, with $v \approx 10^{-3} \ c$. With detector recoil triggers being $\sim \mathcal{O}({\rm keV})$ or larger, the mass of the DM that can be detected is $\sim \mathcal{O}({\rm MeV})$. In order to detect low mass DM particles, we can take any of the following measures. The obvious one is to lower the detector recoil trigger. This involves
finding new detector materials and building new detectors. A lot of work has been undertaken on this
front, notably the use of Superconducting Nanowires to build a device with the threshold energy of $\sim \mathcal{O}({\rm eV})$ \cite{Hochberg:2019cyy, Hochberg:2021yud}. We, however, focus on a strategy that allows us to use existing detector data to put exclusion
limits on low mass DM, viz. by boosting DM particles in the galactic halo using
cosmic ray electrons to relativistic speeds, so that even very low mass DM particles can trigger
the detector.

In this paper, we considered the effect of such a boost as well as the effect of the Lorentz
structure of the couplings, which has been missing in most of the literature till now. DM 
particles can interact with SM electrons via a `dark' mediator, which is charged under both
the DM gauge group and the SM electroweak gauge group. We considered mediators of two
kinds - vector and scalar. For each of the cases, we explored the effect when the 
mediator is very heavy or when the mediator is very light, using data from {\xenon} and 
{\superkam} ({\superk}).

Boosts due to cosmic ray electrons drastically change the DM flux as seen on Earth. The effect
though is quite different for different DM masses as well as for different mediator masses. For 
light mediators, the boosted flux is suppressed below the constant cross-section flux for relatively heavier DM masses, while it is raised above that level for light DM masses. This is 
very different for heavy mediators, for which the DM flux is augmented above the constant
cross-section case for all masses. This behaviour is largely independent of the nature of the
mediator, though there are some numerical differences in the scalar and vector 
cases. This behaviour, in turn, leads us to expect that the energy dependence of the cross-section can provide stronger bounds for lighter DM in the light mediator case, while providing 
stronger bounds for a large range of DM masses in the heavy mediator case. Our analysis 
meets this expectation. 

The two experiments whose dataset  we use differ in two fundamental aspects. While {\superk}
has a much larger number of target electrons (as can be seen by the different values of $\aleph$ 
used in our analysis), {\xenon} has a much smaller trigger energy. The live-time for the dataset
from {\superk} is also longer than for the dataset from {\xenon}. We find that for the light mediator
case, {\xenon} gives stronger bounds on both the cross-section and the electron-mediator 
couplings compared to {\superk}, while for the heavy mediator case, 
the reverse is true. 

The exclusion bounds on the cross-section obtained from our analysis, for the light mediator case, is competitive with that
obtained by the authors of Refs.~\cite{Hochberg:2019cyy, Hochberg:2021yud} using their prototype superconducting nanowire single photon detector (SNSPD) to detect non-relativistic DM
particles for masses above $\sim 1\ {\rm MeV}$. Our bounds extend much further in the 
lower mass regions, however, and are also stronger in the heavy mediator case. Of course, the projected limits using novel materials like NbN and Al are much 
stronger than their current observed limits or ours. 

In the analysis presented here, we tried to calculate the effect of both, boosts for DM particles
and the Lorentz structure of the operators involved. We find that both effects modify
the bounds from the existing constant cross-section case. We also perform a preliminary investigation of the attenuation of DM particles. A more rigorous analysis is in progress and will be 
presented in a future work.

\section*{Acknowledgements}
D.G. acknowledges support through the Ramanujan Fellowship and MATRICS Grant of the
Department of Science and Technology, Government of India. D.B.  acknowledges  financial  
support  through  the National   Postdoctoral   Fellowship   (NPDF),   SERB,  PDF/2021/002206. 
Work of A.G. is supported by the National Research Foundation of Korea 
(NRF-2019R1C1C1005073). D.S. has received funding from the European Union’s Horizon 2020 
research and innovation programme under grant agreement No 101002846, ERC CoG 
“CosmoChart”. The authors also thank Arka Banerjee and Susmita Adhikari for their valuable comments on the Tremaine-Gunn bound. We also thank Robert McGehee for his valuable comments on the pre-print.

\newpage
\onecolumngrid

\appendix

\section{The Ionization Form Factor}
\label{appendix:A}
The cross section for the scattering process $\chi(p) +e(k) \rightarrow \chi(p')+e(k')$ is given by 
\begin{eqnarray}
d\sigma = \frac{|\mathcal{M}|^2}{v_{\chi e}}\frac{1}{64 \pi^2 E_\chi E'_\chi E_e E'_e} \frac{1}{(2\pi)^3} \delta(\Delta E_\chi - \Delta E_e) f_{i\rightarrow k'}(\vec{q})  d^3\vec{q} d^3 \vec{k}'
\label{eq:atomic_formfac}
\end{eqnarray}

where the atomic form factor $f_{i\rightarrow k'}(\vec{q})$ takes care of the initial and final states of the electron. Eqn.~\ref{eq:atomic_formfac} can be recast \cite{Cao:2020bwd} to take the form of Eqn.~\ref{eq:sigmav_ion}, with the ionization form factor $f_{\text{ion}}^{nl} (k',q)$ defined as 

\begin{eqnarray}
|f_{\text{ion}}^{nl} (k',q)|^2 = \frac{2k'^3}{(2 \pi)^3} \sum_{\text{deg}} |f_{i\rightarrow k'}(\vec{q})|^2
\end{eqnarray}

If the initial and final states are free, then this factor reduces to $f_{i\rightarrow k'}(\vec{q})=(2 \pi)^3 \delta^3 (\vec{k} - \vec{k}'+\vec{q})$, which is the case for {\sc{Super-K}}. For {\xenon}, after ionisation, the electron is a free particle, while for the initial state, the contributing electronic orbitals of Xenon are $(5p^6,5s^2,4d^{10},4p^6,4s^2)$. The momentum of the final state is given by $k'=\sqrt{2m_e E_R}$. The expression for the ionization form factor is given by the following \cite{Essig:2011nj,Cao:2020bwd}

\begin{eqnarray}
|f_{\text{ion}}^{nl} (k',q)|^2 = \frac{(2l+1) k'^2}{4\pi^3 q}\int_{|k'-q|} ^{|k'+q|} |\chi _{nl} (k)|^2 k dk
\end{eqnarray}

where the radial wave function in momentum space $\chi_{nl}(k)$ can be expressed as a linear combination of the Slater-type orbitals \cite{Bunge:1993jsz,Kopp:2009et,Cao:2020bwd}, which results in the following expression

\begin{eqnarray}
\chi_{nl}(k)= \sum_j C_{nlj} 2^{n_{lj} -l} \left(\frac{2\pi a_0}{Z_{lj}}\right)^{3/2} \left( \frac{ipa_0}{Z_{lj}} \right)^l \frac{\Gamma(n_{lj}+l+2)}{\Gamma(l+\frac{3}{2}) \sqrt{(2n_{lj})!}} 
\nonumber \\
\times {}_2F_1 \left[\frac{1}{2} (n_{lj} +l+2), \frac{1}{2}(n_{lj}+l+3), l+\frac{3}{2},- \left(\frac{pa_0}{Z_{lj}} \right)^2 \right]
\end{eqnarray}

Here, ${}_2F_1$ denotes the hypergeometric function, $a_0$ is the Bohr radius, and the coefficients $C_{nlj}, Z_{lj}$ and $n_{lj}$ are taken from Ref.~\cite{Bunge:1993jsz}.

\bibliography{reference}

\end{document}